# Nonlocal Interferometry - Beyond Bell's Inequality

**J.D. Franson**
*University of Maryland, Baltimore County, Baltimore, MD 21250*
*jfranson@umbc.edu*

**Abstract:** This paper presents a tutorial review of nonlocal interferometry. The role of the Feynman propagator in optical coherence is also discussed, including the possibility of generating optical coherence and entanglement outside of the light cone.
©2007 Optical Society of America

## 1. Introduction

The phase associated with a quantum system can exhibit a variety of nonclassical properties, such as the Aharonov-Bohm effect [1] or Berry's geometric phase [2]. Nonlocal interferometry is another example of the nonclassical phase properties of quantum systems. In addition to providing insight into the nature of optical coherence, nonlocal interferometry has a number of practical applications as well. This paper is intended to provide a tutorial review of nonlocal interferometry.

In addition to reviewing earlier work on nonlocal interferometry, some recent results related to the role of the Feynman propagator [3] in optical coherence will also be discussed. The Feynman propagator has nonzero values outside of the forward light cone. That does not allow messages to be transmitted faster than the speed of light, but it does allow optical coherence and entanglement to be generated outside the forward light cone [4]. The possible impact of effects of this kind on the theory of optical coherence will be considered.

As a graduate student at Caltech, I had the opportunity to take Richard Feynman's class on quantum mechanics, which made no mention of Bell's inequality. At some point, one of the students asked Feynman if he would explain Bell's inequality. Feynman's reply was "There is nothing to it – I will explain it all later". But he never did. That was the only time I heard anyone ask Feynman a question that he didn't immediately answer in a very clear way, and this anomaly was one of the reasons that I became interested in Bell's inequality and eventually nonlocal interferometry.

## 2. Nonlocal phase measurements

Perhaps the best place to begin is with the nonlocal phase measurements [5] illustrated in Fig. 1. A number state containing a large number $N$ of photons is assumed to be incident on a 50/50 beam splitter. After the two beams are separated by a large distance, a small fraction of the energy in beam A is further split off and its phase is measured by comparison with a local oscillator. It can be shown that the state of the remaining photons in beam A will then collapse to a coherent state with the measured value of the phase. More interestingly, the state of beam B will also collapse to a coherent state with the same phase, aside from any difference due to its propagation. Klaus Molmer has also discussed some similar situations [6].

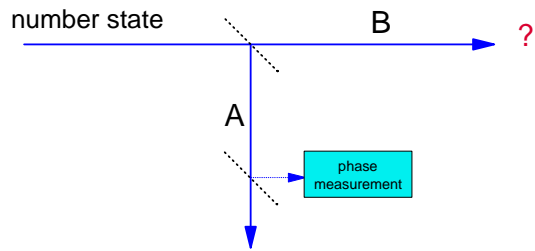

Fig. 1. Nonlocal phase measurements [5] in which an incident photon number state is split into two distant beams A and B. A measurement of the phase of beam A will reduce beam B to a coherent state with the same phase.

This is an example of the Einstein-Podolsky-Rosen (EPR) paradox. According to local realism, the phase of the number state must have been determined all along even if we don't know what it is. What does this say about the nature of optical coherence? In the quantum-mechanical interpretation, the beam in path B was totally incoherent before the measurement and then it suddenly becomes totally coherent immediately after the measurement.

This example does not violate Bell's inequality and the results of such an experiment could not distinguish between the local realistic interpretation and the quantum-mechanical interpretation. Nonlocal interference experiments of the kind described in the next section can violate Bell's inequality and they can distinguish between the two types of theories.

### 3. Energy-time entanglement

Fig. 2 illustrates a nonlocal interference experiment [7] in which two distant interferometers each have a long path L and a short path S. A nonclassical source emits a pair of entangled photons that propagate towards the two interferometers. In classical optics, the output of each interferometer would be determined by the local phase difference between the two paths through the corresponding interferometer. But in nonlocal interferometry of this kind, the output of both interferometers can depend on the values of both phase differences. It can be shown that the coincidence rate $R_C$ for detecting a photon at the same time in both of the horizontal output ports, for example, is given by

$$R_C = \alpha \cos^2\left[(\phi_1 + \phi_2)/2\right]. \qquad (1)$$

These results violate Bell's inequality and they are inconsistent with any local realistic theory. Roughly speaking, the value of both phase settings would have to be known at the location of each interferometer in order to reproduce these results classically. If the settings of the phases were changed rapidly enough, information would have to be transmitted faster than the speed of light in order to reproduce these kinds of results in any classical theory.

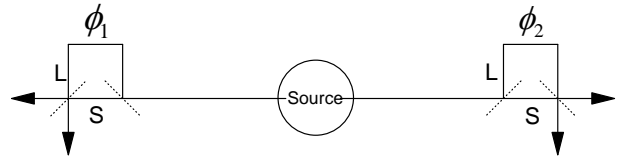

Fig. 2. Nonlocal interferometer [7] in which a pair of energy-time entangled photons are incident on two interferometers with a long path L and a shorter path S.

Interference effects of this kind require that the photons be entangled in energy and time [7]. Entanglement of that kind can be produced using an atomic cascade in a three-level atom, as illustrated in Fig. 3. It is assumed that the transition time $\tau_1$ for the first atomic transition is much longer than the transition time $\tau_2$ for the second transition. Since $\tau_1$ is very long, the total energy of the system is well defined and the sum of the energies of the two photons must equal the change $\Delta E_A$ in the energy of the atom:

$$\hbar\omega_1 + \hbar\omega_2 = \Delta E_A. \qquad (2)$$

As a result, the state of the system is entangled in energy and it has the approximate form

$$|\psi\rangle \sim \int d\omega_1 \hat{a}_1^\dagger(\omega_1)\hat{a}_2^\dagger(\Delta E_A/\hbar - \omega_1)|0\rangle \qquad (3)$$

where the operators $\hat{a}_1^\dagger(\omega_1)$ and $\hat{a}_2^\dagger(\omega_2)$ create photons of frequency $\omega_1$ and $\omega_2$ in the two paths. The frequencies of the two photons are anti-correlated in an entangled state of this kind.

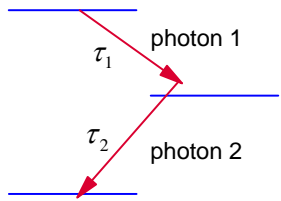

Fig. 3. An atomic cascade with $\tau_1 \gg \tau_2$, which produces pairs of energy-time entangled photons.

On the other hand, the two photons must have been emitted at the same time in the limit where $\tau_2$ is very short. This implies that the state of the system could also be written approximately as

$$|\psi\rangle \sim \int dt \hat{E}_1^{(-)}(t)\hat{E}_2^{(-)}(t)|0\rangle. \qquad (4)$$

This corresponds to a coherent superposition of times at which both photons may have been emitted and the two photons are thus entangled in the time domain as well. Equations (3) and (4) are related to each other by a Fourier transform and the state is said to be energy-time entangled. Similar states can also be prepared using parametric down-conversion.

The operation of the interferometer shown in Fig. 2 can now be understood from the fact that the two photons are emitted at the same time. Since we only accept events in which the photons arrive at the same time, both photons must have traveled the same distance. Quantum interference between the $L_1L_2$ and $S_1S_2$ probability amplitudes give rise to the coincidence rate of Eq. (1).

These effects seem straightforward now but they were highly controversial at the time. Most of the skepticism was due to the fact that the ordinary (first order) coherence length of the photons is assumed to be much less than the difference in the path lengths $L-S$. In classical optics, there would be no interference under conditions of that kind. This situation is illustrated in Fig. 4, where a short classical pulse whose length is equal to the coherence length is split apart when it propagates through one of the interferometers, producing no overlap and no interference. My view of the situation was that the wave packets actually had a much longer length equal to $c\tau_1$, as illustrated by the red curve in Fig. 4, and that the shorter wave packets do not exist until the position of one of the photons is measured. In any event, the second-order coherence length can be much larger than the first-order coherence length, which allows the operation of interferometers of this kind with relatively large path length differences.

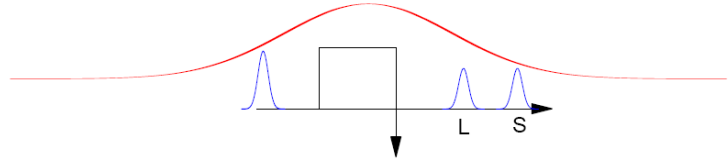

Fig. 4. Propagation of a short classical pulse through an unbalanced interferometer, which does not give any interference effects. The second-order coherence length corresponds to a much longer wave packet (upper curve), which explains why quantum interference can occur in the nonlocal interferometer of Fig. 3 even though the usual coherence length is much shorter than the path difference.

The fact that the first-order coherence length is very short can be used to derive [8] an inequality that must be satisfied by any classical field theory:

$$v \leq \frac{R_{C0}(\Delta t)}{R_{C0}(0) + R_{C0}(\Delta t)}. \qquad (5)$$

Here $R_{C0}(\Delta t)$ is the coincidence counting rate that would be obtained with no interferometers in place for a time delay of $\Delta t$ and $v$ is the visibility of the two-photon interference fringes. Unlike Bell's inequality (which requires $v > 71\%$), this inequality can be violated for visibilities that are much less than $50\%$.

The first experimental demonstration of nonlocal interferometry of this kind was performed by Kwiat, Vareka, Hong, Nathel, and Chiao using two parallel paths in a single interferometer [9]. The first demonstration using two separated interferometers was achieved by Ou, Zou, Wang, and Mandel [10]. I performed an early experiment over what was then a large path length of 100 m in free space [11]. Martienssen's group was the first to achieve visibilities high enough to violate Bell's inequality [12], and Rarity and Tapster [13] were the first to couple one of the photons into an optical fiber. A number of subsequent experiments were performed with increasing visibilities [14].

The separation of two particles is an essential part of nonlocality, and it seems natural to ask what happens as the separation between the interferometers is increased. The separation should make no difference according to quantum mechanics, but it is conceivable that some kind of intrinsic decoherence mechanism might reduce the visibilities over sufficiently large distances. Interferometers of this kind are ideal for long-distance experiments, since the visibility is not affected by changes in the state of polarization of the photons nor by any phase shifts that occur along the path of propagation. Gisin's group has done a series of experiments in optical fibers resulting in violations of Bell's inequalities over distances greater than 50 km.

## 4. Other types of interferometers

The well known Hong-Ou-Mandel interferometer [15] is illustrated in Fig. 5. Here two identical photons are assumed to be incident at the same time on a 50-50 beam splitter. There is a probability amplitude for both photons to be reflected, and this is equal and opposite to the probability amplitude for both photons to be transmitted. As a result, the photons will not propagate into two different output ports of the beam splitter if they arrive at the same time. A measurement of the coincidence counting rate in the two output ports as a function of the relative time delay in the arrival of the photons is shown in Fig. 6. It can be seen that there is large dip in the coincidence counting rate at zero time delay. This particular data was collected in our lab and it corresponds to a visibility of greater than 99%. A similar effect was also demonstrated by Shih and Alley [16].

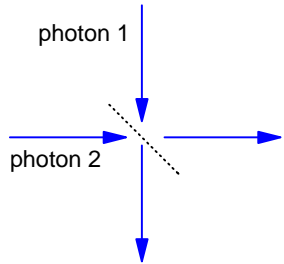

Fig. 5. The Hong-Ou-Mandel interferometer in which two indistinguishable photons are incident on a single beam splitter.

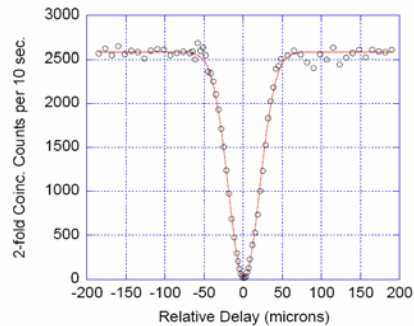

Fig. 6. Dip in the coincidence counting rate obtained in a Hong-Ou-Mandel interferometer when the arrival time of one photon is delayed with respect to the other photon.

The Hong-Ou-Mandel interferometer can be viewed as a local effect, since the two photons interfere on a single beam splitter, and it does not violate Bell's inequality. Nevertheless, it is a very useful device for measuring small changes in path lengths or propagation times, and it is a very convenient way to measure the degree of indistinguishability of two photons.

Another interferometer that does violate Bell's inequality was proposed by Horne, Shimony, and Zeilinger [17]. As illustrated in Fig. 7, this interferometer is based on entanglement between the paths taken by a pair of photons. Both photons may propagate in paths A and B, for example, or both photons may propagate in paths C and D. A superposition of these two probability amplitudes corresponds to an entangled state. Interference fringes can be obtained by combining the two sets of paths with two beam splitters. Rarity and Tapster [18] were the first to experimentally demonstrate nonlocal interferometry based on path entanglement. Scully and Druhl [19] independently proposed a similar effect.

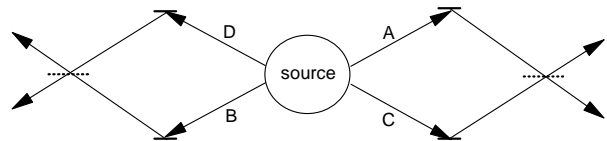

Fig. 7. Nonlocal interferometer based on path entanglement [17].

Gisin's group [20] has demonstrated a related form of entanglement based on the use of time bins.  This is similar to energy-time entanglement except that the energy entanglement has been removed as illustrated in Fig. 8.  Here a short laser pulse is incident on an unbalanced interferometer to produce two successive output pulses.  The two pulses are then passed through a down-conversion crystal to produce a pair of photons.  As shown in the figure, there is some probability amplitude to produce a pair from the first pulse and an equal probability to produce a pair from the second pulse.  The time-entangled photons can then be routed to two distant interferometers as in Fig. 2 to produce nonlocal interference effects and violations of Bell's inequality.  This approach has some practical advantages, such as the use of gated detectors, for example.  On the other hand, it does not allow the use of nonlocal dispersion cancellation, as described below.

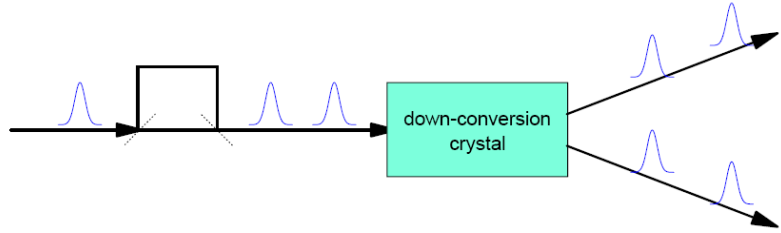

Fig. 8.  Source of time-bin entangled photons.  A single laser pulse is split into two pulses in an unbalanced interferometer and passed through a parametric down-conversion crystal.  This produces an equal probability amplitude for a pair of secondary photons to have been emitted at two different times.

## 5. Nonlocal cancellation of dispersion

Nonlocal cancellation of dispersion is another example of a quantum interference effect.  In classical optics, the dispersion of a pulse of light in one medium is independent of the properties of a distant dispersive medium, as illustrated in Fig. 9.  In quantum optics, however, the dispersion present in one medium can cancel the dispersion of a distant medium if a pair of energy-time entangled photons propagate through the two distant media [21].

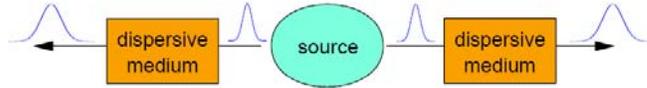

Fig. 9.  Propagation of two classical pulses of light through two dispersive media.  In quantum optics, the dispersion in one medium can cancel the dispersion in the other medium [21].

Consider two distant media whose second-order dispersion coefficients are equal and opposite:

$$\frac{d^2 k_1}{d\omega_1^2} = -\frac{d^2 k_2}{d\omega_2^2}. \qquad (6)$$

From Eq. (3), the frequencies of the two photons are anti-correlated, so that the dispersive phase shift due to the $\hat{a}_1^\dagger(\omega_1)$ term will be equal and opposite to that from the $\hat{a}_2^\dagger(\omega_2)$ term.  As a result, there will be no net phase shift and the state of the system will be unaltered by the dispersive media: the two photons will remain coincident after they have propagated through the two media.  These results can also be shown to violate a classical inequality and there is no classical analogy for this effect [21].

Although the dispersion conditions of Eq. (6) are difficult to satisfy in general, optical fibers near the telecommunications band at 1.55 $\mu m$ do have this property.  The second-order dispersion coefficient is negative below a certain wavelength and positive above it.  Gisin's group [22] has used this feature to demonstrate nonlocal dispersion cancellation over large distances in telecommunications fibers using broadband entangled photon pairs from parametric down-conversion.  This is an important practical consideration in quantum key distribution systems based on entangled photon pairs and nonlocal interferometry.

Steinberg, Kwiat, and Chiao [23] independently proposed and demonstrated a different kind of dispersion cancellation illustrated in Fig. 10. Here a dispersive element is located in one path of a Hong-Ou-Mandel interferometer. It can be shown that the dip in the coincidence curve of Fig. 6 is unaffected by the dispersive element. This is a higher-order quantum interference effect as before, but it can be viewed as a local phenomenon since the two photons interfere at a single beam splitter instead of in two distant media. As a result, this form of dispersion cancellation does have a classical analogy [24] (with reduced visibility). Nevertheless, this is an extremely useful technique with possible practical applications in microscopy and high-resolution timing measurements. Steve Harris has recently shown that dispersion cancellation effects of both kinds can be useful in generating half-cycle biphotons [25].

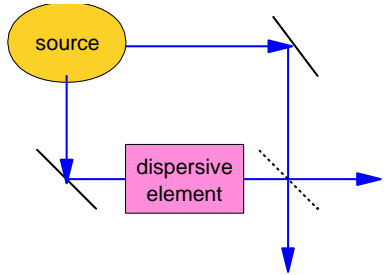

Fig. 10. Cancellation of dispersion in a Hong-Ou-Mandel interferometer, as proposed and demonstrated by Steinberg, Kwiat, and Chiao [23].

## 6. Quantum interference using other kinds of particles

My original paper [7] on nonlocal interferometry actually used general field operators $\hat{\psi}^\dagger(r,t)$ instead of the electric field operator, and the entangled state of the system can be written in the approximate form

$$|\psi\rangle \sim \int dt\, \hat{\psi}^\dagger(r_1,t)\hat{\psi}^\dagger(r_2,t)|0\rangle. \qquad (7)$$

In principle, all of the same effects can be obtained using other kinds of particles, or even two different kinds of particles such as an electron and a neutron. At the time, it did not seem likely that nonlocal interferometry experiments would ever be performed using particles other than photons, but there are now two situations in which that is close to being the case.

Gisin's group [26] has demonstrated second-order coherence using energy-time entangled surface plasmons. A pair of entangled photons is converted to a pair of surface plasmons, which propagate some distance before they are converted back into a pair of photons and passed through two interferometers as in Fig. 2. Surface plasmons are collective modes of a macroscopic number of electrons, and these experiments clearly show that systems of that kind can exist in energy-time entangled states.

Our group has recently been considering the possibility of violating Bell's inequality in nonlocal interferometry experiments that make use of entangled photon holes [27,28], which are somewhat analogous to the electron holes of semiconductor theory. Entangled photon holes can best be understood by comparison with parametric down-conversion, as illustrated in Fig. 11. In parametric down-conversion, the two output beams are originally empty (in the vacuum state) and pairs of entangled photons are created at the same time, with a coherent superposition of such times. In the case of entangled photon holes, two weak laser beams are assumed to be incident on a two-photon absorbing medium. Here the two output beams initially have a uniform probability amplitude, and the absorption of pairs of photons create "holes" in the otherwise uniform background. It can be shown that the entangled photon holes can violate Bell's inequality if they are passed through two distant interferometers, as in Fig. 2. We have experimentally demonstrated the entangled photon holes [28] but an experiment to violate Bell's inequality is still in progress.

## 7. The Feynman propagator

Nonlocal interferometry can be used to violate Bell's inequality, which is inconsistent with any local realistic (hidden variable) theory. One interpretation of such experiments is that the actual state of the system is unknown until we measure it, which does not imply any need for information to propagate faster than the speed of light. I would now like to consider another type of phenomenon that goes beyond Bell's inequality in the sense that it does require the propagation of virtual particles outside of the forward light cone. In particular, I recently showed that optical coherence and entanglement could be generated between two distant locations in less time than it would take for light to travel between them [4].

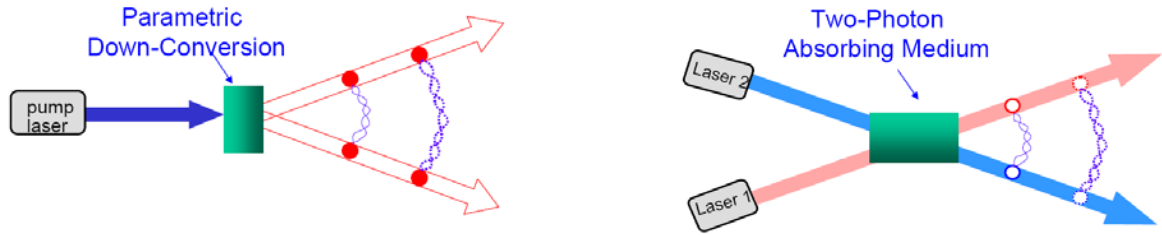

Fig. 11. Comparison of the generation of entangled photons using parametric down-conversion (left-hand side) with the generation of entangled photon holes using a two-photon absorbing medium (right-hand side).

The probability amplitude to emit a photon at $\mathbf{r}_1, t_1$ and then annihilate it at $\mathbf{r}_2, t_2$ is proportional to the Feynman propagator $D_F(\mathbf{x}_2 - \mathbf{x}_1, t_2 - t_1)$. The Feynman propagator [3] has nonzero values outside of the forward light cone, as can be seen from the plot in Fig. 12. That does not allow messages to be transmitted faster than the speed of light, but it does allow optical coherence, entanglement, and mutual information to be generated outside of the light cone. Several possible interpretations of these results will be discussed below.

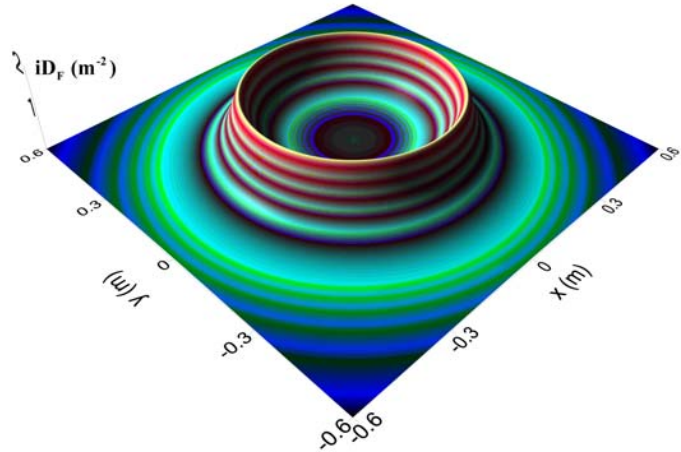

Fig. 12. A plot of the Feynman propagator in the xy-plane. It can be seen that the Feynman propagator has nonzero values outside of the forward light cone (the yellow ring).

Feynman did discuss the Feynman propagator in his quantum mechanics class. In his published comments [29], he states that "Possible trajectories are not limited to regions within the light cone" but "in reality, *not much* of the space outside of the light cone is accessible." The italics are mine: to me, the question was always whether or not there are any effects outside of the light cone, regardless of their magnitude or extent. In his classroom lectures, Feynman simply said that causality does not apply over sufficiently short distances, such as the Compton wavelength of the electron. But the scale of Fig. 12 is on the order of unity in MKS units along all three axes, and it is not so apparent that the effects are confined to microscopic distances.

The effects of the Feynman propagator can be understood by considering two distant atoms, as shown in Fig. 13. Atom 1 is assumed to be initially excited while atom 2 is initially in its ground state. The question is whether or not atom 1 can emit a photon and make a transition to its ground state while atom 2 absorbs the photon and makes a transition to its excited state within some time interval $\Delta t$. This problem was first considered by Fermi [30], who made several unwarranted assumptions and concluded that there were no effects at all outside of the forward light cone. Several authors considered the problem in more detail and concluded that the state of atom 2 could be affected outside the light cone in apparent violation of causality. It was later shown that the expectation value of the state of atom 2 could not change outside the light cone, but that correlations could be generated [31]. Ref. [4] generalizes this to show that maximal entanglement and mutual information can be generated as well.

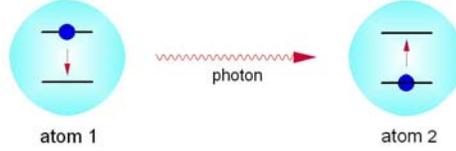

Fig. 13. Two atoms, one initially in its excited state and the other in its ground state. Atom 1 can emit a photon and make a transition to its ground state while atom 2 can absorb the photon and make a transition to its excited state. The probability amplitude for this process is proportional to the Feynman propagator, which is nonzero outside the forward light cone.

I recently learned that B. Reznik et al. [32] obtained similar results for a scalar field. They interpreted this as a transfer of pre-existing entanglement in the quantum vacuum to the atoms. The existence of entanglement in the quantum vacuum is controversial [33] and I do not agree with that interpretation, but it does raise the question as to what is the role of the quantum vacuum in situations of this kind. Several possible interpretations will be discussed below.

The role of the Feynman propagator can be seen by using second-order perturbation theory to calculate the change $|\Delta \psi(\Delta t)\rangle$ in the state vector:

$$|\Delta \psi(\Delta t)\rangle = \frac{1}{(i\hbar)^2} \int_0^{\Delta t} dt' \int_0^{t'} dt'' \hat{H}'(t')\hat{H}'(t'') |\psi_0\rangle. \tag{8}$$

Here $\hat{H}'(t)$ is the usual interaction Hamiltonian in the interaction picture. If we make the dipole approximation, the atomic matrix elements become constants and the probability amplitude $b$ to be in the state $|G_1\rangle|E_2\rangle|0\rangle$ is given by

$$b = \frac{1}{(i\hbar)^2}\left(\frac{edE_A}{\hbar c}\right)^2 \int_0^{\Delta t} dt' \int_0^{t'} dt'' e^{-iE_A(t''-t')/\hbar} \langle 0|\hat{A}_x(\mathbf{x_2},t')\hat{A}_x(\mathbf{x_1},t'')|0\rangle. \tag{9}$$

Here $d$ is the dipole moment, $E_A$ is the energy difference between the two atomic states, and $\hat{A}_x(\mathbf{x},t)$ is the electric field operator along the direction of $d$.

The matrix elements of the field operators can be evaluated using commutator techniques:

$$\begin{aligned}
\langle 0|\hat{A}_x(\mathbf{x_2},t')\hat{A}_x(\mathbf{x_1},t'')|0\rangle &= \langle 0|\hat{A}_x^{(+)}(\mathbf{x_2},t')\hat{A}_x^{(-)}(\mathbf{x_1},t'')|0\rangle \\
&= \langle 0|\hat{A}_x^{(-)}(\mathbf{x_1},t'')\hat{A}_x^{(+)}(\mathbf{x_2},t') + [\hat{A}_x^{(+)}(\mathbf{x_2},t'),\hat{A}_x^{(-)}(\mathbf{x_1},t'')]|0\rangle \\
&= \langle 0|[\hat{A}_x^{(+)}(\mathbf{x_2},t'),\hat{A}_x^{(-)}(\mathbf{x_1},t'')]|0\rangle \\
&= -ic\hbar D_F(\mathbf{x_2}-\mathbf{x_1},t'-t'').
\end{aligned} \tag{10}$$

The commutator is proportional to the Feynman propagator in the Lorentz gauge. For a massless particle such as a photon

$$D_F(x_2 - x_1) = -\frac{1}{4i\pi^2} \frac{1}{|\mathbf{x_2}-\mathbf{x_1}|^2 - c^2(t'-t'')^2 - i\varepsilon}. \tag{11}$$

In the limit where the distance r is much larger than $c\Delta t$, the Feynman propagator is inversely proportional to $r^2$ and the final state of the system is given by

$$|\psi\rangle = a|E_1\rangle|G_2\rangle + b|G_1\rangle|E_2\rangle + \gamma|\phi_\perp\rangle \quad (12)$$

where

$$b = -\frac{\alpha}{4\pi^2}\frac{d^2}{r^2}\left(i\omega_A\Delta t + 1 - e^{i\omega_A\Delta t}\right). \quad (13)$$

Here $|\phi_\perp\rangle$ is orthogonal to the other two terms and it reflects the possibility that atom 1 may emit a photon that is not absorbed by atom 2, while $\alpha$ is the fine structure constant and $\gamma$ is a constant of no interest.

## 8. Generation of entanglement outside the light cone

It can be shown [4] that Eq. (12) corresponds to an entangled state in the sense that

$$|\psi\rangle \neq |\Psi_1\rangle|\Psi_2\rangle. \quad (14)$$

Here $|\Psi_1\rangle$ corresponds to the most general state of atom 1 and the field, and similarly for $|\Psi_2\rangle$. This is an entangled state of the atoms and the field, but a maximally entangled state of the atoms alone can be obtained using post-selection as illustrated in Fig. 14. First an array of detectors is used to post-select events in which no photons remain. The state of the system is then reduced to

$$|\psi\rangle = a'|E_1\rangle|G_2\rangle + b'|G_1\rangle|E_2\rangle \quad (15)$$

which is an entangled state. In order to produce a maximally entangled state, we would like to be able to make the coefficient $b'$ equal to $a'$. This can be done by applying a suitable laser pulse to atom 2 to transfer some of the probability amplitude from state $|G_2\rangle$ to some other state $|F_2\rangle$. A measurement is then performed to determine if atom 2 is in state $|F_2\rangle$; if not, the system will be projected into a maximally entangled state with $a' = b'$ for the proper choice of laser pulse intensity. It should be noted that the entire post-selection process can be completed and recorded by the detectors in a time interval that is comparable to $\Delta t$, so that the entangled state is created in less time than it would take for light to travel between the two atoms. Some applications would, however, require a much longer time in order to determine which events succeeded.

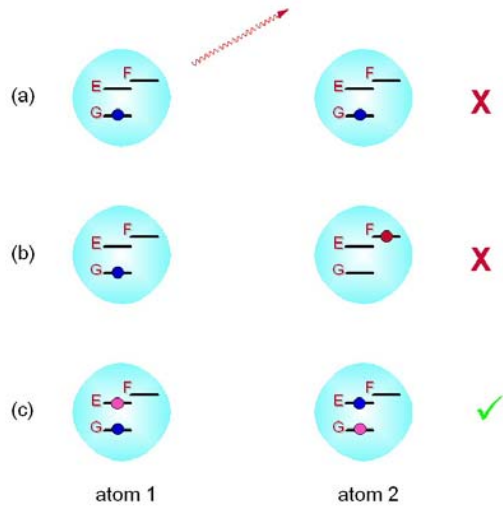

Fig. 14. Use of post-selection to produce maximally entangled pairs of photons outside of the light cone.

The process described above could, in principle, be repeated in parallel using multiple pairs of atoms to generate multiple sets of entangled atoms. We can consider the classical mutual information generated in this way if we simply measure the states of the atoms and take the ground state to represent the bit 0 and the excited state to represent the bit 1, for example. It can be shown that this process leads to the generation of classical mutual information outside of the light cone, either with or without post-selection. In principle, the generation of entanglement outside of the light cone would allow new types of quantum information protocols, such as a quantum time capsule [4]. As a practical matter, the rate at which the entanglement is generated would be too small to be of any real use.

The analysis described above corresponds to the bare-state basis where the atoms are assumed to initially be in their ground state or excited state without the presence of any photons. The problem can also be analyzed in the dressed-state basis which includes the presence of virtual photons in the initial state, as illustrated in Fig. 15. It can be shown that the generation of entanglement outside of the forward light cone is independent of any virtual photons or entanglement that may have existed prior to the interaction time $\Delta t$. As a result, the effects are the same in the dressed state basis of Fig. (15a) and (15b) as it is in the bare-state basis shown in Figs. (15c) and (15d).

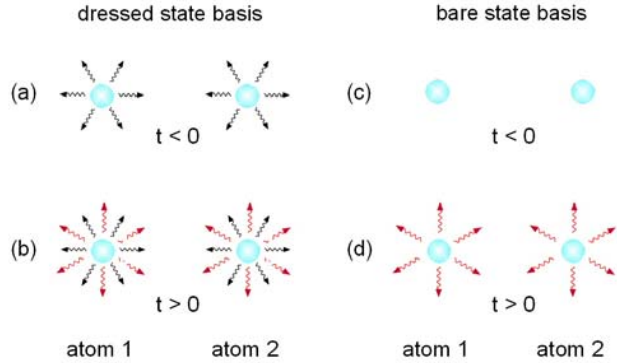

Fig. 15. Comparison of the interaction of two atoms in the dressed state basis and the bare state basis. In the dressed-state basis, the initial state (a) already contains virtual photons. During the interaction time, additional virtual photons are emitted, as illustrated by the red arrows (b). The results obtained in the dressed state basis are the same to lowest order as those obtained in the bare state basis, where there are no virtual photons in the initial state (c) and only those emitted during the interaction time need to be considered (d).

## 9. Optical coherence and the Feynman propagator

The usual expression [34] for the probability of detecting a photon is given by

$$P_d = \eta \langle \hat{E}^{(-)}(\mathbf{x},t)\hat{E}^{(+)}(\mathbf{x},t) \rangle \qquad (16)$$

where $\eta$ is a constant related to the detection efficiency. It is straightforward to use the Feynman propagator to show that this expression would lead to nonzero detection probabilities outside of the light cone, which would be a violation of causality. It has been known for a long time that this difficulty is due to the use of the rotating wave approximation (RWA) and the situation was discussed in some of Glauber's later papers. Roughly speaking, the RWA approximation neglects the probability that atom 2 can make a transition from its ground state to its excited state while emitting a photon of its own, since energy is not strictly conserved over a finite time interval $\Delta t$. It would not be difficult to generalize Eq. (16) to avoid these difficulties.

The situation is more complicated, however, if we consider expressions for the higher-order optical coherence, such as

$$g^{(2)}(\mathbf{x_1},t_1;\mathbf{x_2},t_2) = \frac{\langle \hat{E}^{(-)}(\mathbf{x_1},t_1)\hat{E}^{(-)}(\mathbf{x_2},t_2)\hat{E}^{(+)}(\mathbf{x_2},t_2)\hat{E}^{(+)}(\mathbf{x_1},t_1) \rangle}{\langle \hat{E}^{(-)}(\mathbf{x_1},t_1)\hat{E}^{(+)}(\mathbf{x_1},t_1) \rangle \langle \hat{E}^{(-)}(\mathbf{x_2},t_2)\hat{E}^{(+)}(\mathbf{x_2},t_2) \rangle}. \qquad (17)$$

Although this equation is a definition, its structure is similar to Eq. (16) and the RWA approximation would be required in order to relate this equation to experimentally observable detection events. In any event, it can be shown that the use of this equation would also allow messages to be transmitted faster than the speed of light. Once again, this result is not physical due to the use of the RWA approximation, unlike the correlations and entanglement discussed in the preceding sections.

Approximate solutions to this problem have been discussed [35,36]. There is some reason to suspect that the concept of optical coherence may be inherently approximate, since $\langle \hat{E}(r,t)\hat{E}(r,t)\rangle$ at a point is divergent. The divergence can be eliminated using normal ordering as in Eqs. (16) and (17), but that would appear to bring in the RWA approximation in some way. Alternatively, we could average the fields over some finite regions of space, but the averaging process would also be approximate and would depend on the size of the regions. This is an interesting issue for further study.

## 10. Discussion and conclusions

In my opinion, there is not much point is debating different interpretations that all correspond to the same basic physics. Nevertheless, one might inquire as to what is the "correct" interpretation of the generation of entanglement and mutual information outside of the forward light cone. Three possible interpretations are briefly discussed..

One interpretation is to assume as usual that electromagnetic interactions are produced by the exchange of virtual photons that carry energy and momentum. In that case, the entanglement would be established by the propagation of virtual photons outside the light cone as described by the Feynman propagator. That is what I was taught as a student and I have not heard a better explanation. On the other hand, the photons cannot be directly observed without destroying the effects of interest and one might ask whether or not they are really being exchanged in that case. In addition, the direction of travel of the photons depends on the choice of reference frames [4].

Similar effects for scalar fields have been interpreted by B. Reznik and his colleagues [32,37] as being due to the transfer of entanglement that was already in the quantum vacuum to the atoms. In that case, no superluminal generation of entanglement would be required. Their interpretation is based in part on the Unruh effect, in which an accelerating observer would perceive the quantum vacuum to contain additional particles in an entangled (squeezed) state. But there are no accelerating observers in this problem, and the usual quantum vacuum corresponds to a product state of a large number of harmonic oscillators all in their ground state, which is not an entangled state. The existence of entanglement in the quantum vacuum is controversial [33] and I do not agree with this interpretation.

Even if the quantum vacuum is not entangled, the vacuum fluctuations at two different points may be correlated [38] as illustrated in Fig. 16a. In that case, it may be possible for the local fields to produce changes in the states of the atoms without any transfer of momentum or energy outside of the light cone. But it seems to me that something is missing in this picture, namely the virtual photons depicted in Fig. 16b. From momentum conservation or the form of the electric field operator, the state of the atoms cannot change unless the state of the field is changed, which corresponds to the emission or absorption of virtual photons [4]. The virtual photons presumably carry energy and momentum, and we are left with the first interpretation once again. The effects described here are consistent with zero-point energy and vacuum fluctuations, but there is more to quantum electrodynamics than just the quantum vacuum.

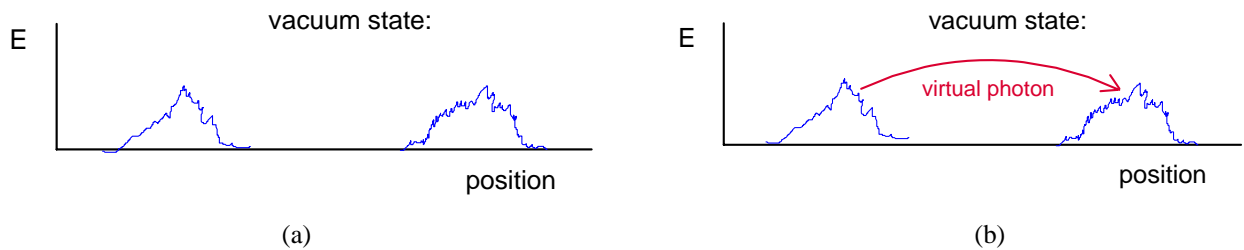

Fig. 16. Possible effects of correlations between the vacuum fluctuations at two different locations on the two atoms of Fig. 13. If the field remains in the vacuum state (a), there can be no change in the state of the atoms. Any change in the state of the atoms requires the emission or absorption of virtual photons (b), which carry energy and momentum outside the light cone.

In summary, nonlocal interference effects lead to violations of Bell's inequality and they may have practical applications as well. The fact that the Feynman propagator has nonzero values outside of the forward light cone does not allow messages to be transmitted faster than the speed of light but it does allow entanglement and mutual information to be generated outside the forward light cone. Regardless of their interpretation, these effects depend on the propagation of virtual photons outside of the light cone and they are superluminal in that sense.

A number of questions remain: Are there any feasible experiments to test effects of this kind outside of the light cone? If an experiment were to be performed, would these effects be observed? And are there any alternative theories that do not have this property? Further research of this kind appears to be a natural extension of the earlier work on Bell's inequality.